\magnification1200
\rightline{KCL-MTH-01-08}
\rightline{hep-th/0104081}
\vskip .5cm
\centerline
{\bf{{$E_{11}$ and M theory. }}}
\vskip 1cm
\centerline{ P. West}
\vskip .2cm
\centerline{Department of Mathematics}
\centerline{King's College, London, UK}

\leftline{\sl Abstract}
We argue that eleven dimensional supergravity can be described  by 
a non-linear realisation based on the group $E_{11}$. 
This requires a formulation of eleven dimensional supergravity in
which the gravitational  degrees of freedom are  described by two
fields which are related by duality. We show the existence of such a
description of gravity. 
\vskip .5cm

\vfill\eject


\medskip 
{\bf {0. Introduction }}
\medskip
There is a widespread belief that all superstring theories are 
manifestations of a single theory that has been called M theory. 
However, little is known for sure about M theory.  
 The conjectured existence of M  theory has
largely relied on the  properties of the  supergravity
theories. In particular,  the unique 
eleven dimensional
supergravity [1],  the maximally supersymmetric ten dimensional
supergravity theory which are the IIA [2,3,4] and IIB  [5]
supergravity theories as well as the type I  supergravity coupled to
the Yang-Mills theory [6]. These theories are essentially determined
by  the type of supersymmetry they possess and hence they are the 
low energy effective actions of the string theories with the
corresponding space-time supersymmetry.  One intriguing feature of
supergravity theories is the occurance of  coset space symmetries
that  control the way the scalars occur in these theories. 
The four dimensional
$N=4$ supergravity theory possess a   SL(2,R)/ U(1) symmetry [9], 
the IIA theory a  SO(1,1)  symmetry [2],  the IIB 
theory a SL(2,R)/ U(1) [5] and the further 
reductions of the eleven dimensional 
 supergravity theory possess cosets based on the exceptional groups  
[10,11,25]. These symmetries have also played an 
important role in string dualities in recent years [12,13].  
The coset construction was extended [14] 
to include the gauge fields of
supergravity theories. This method used generators that were inert
under Lorentz transformations and, as such, it is difficult to extend
to include either gravity or the fermions. However, this construction 
did included the  gauge and scalar fields as well as  their duals and
as a consequence  the  equations of motion for these fields could
be expressed as a generalised self-duality condition.  
\par
Recently [15], it was shown that the entire bosonic section of eleven
dimensional supergravity and the ten dimensional IIA
supergravity theory could be formulated as a non-linear realisation.
This work made use of two old papers [16,17] which formulated 
gravity as a non-linear realisation.  In this way of proceeding 
gravity and the gauge fields appeared on an equal footing. It is very
likely that this construction can be extended to the bosonic sector
of all supergravity theories. Although the extension to include the 
fermions was not given in reference [15] it was realised that this
step implied that eleven dimensional supergravity must be invariant
under a group that included Osp(1/64) as a subgroup. 
\par
The appearance of the above mentioned coset space symmetries in 
dimensional reductions of eleven dimensional supergravity on a torus
can also be viewed as a consequence of supersymmetry in the
dimensionally reduced theory.  However, despite their importance,
there is no really simple explanation of why these symmetries
occur.    It has been shown [18,19] that eleven dimensional
supergravity  does possess an SO(1,2)$\times$SO(16) symmetry,
although the SO(1,10) tangent space symmetry is
no longer apparent in this formulation.  However, it  has been
thought that the exceptional groups found in the dimensional
reductions can not be  symmetries of eleven dimensional
supergravity.  Despite this, it has been noticed that some of the
objects which appear in the reductions do appear naturally in the
unreduced theory [20].   A similar phenomenon occurs in the reduction
of string theories on a torus. For example, if one takes the closed
bosonic string to live on the unique self-dual twenty six dimensional
Lorentzian lattice one finds that it is invariant under the fake
monster algebra [21]. 
\par
It is indeed peculiar that  symmetries should appear in a theory
when they are not present in the  theory before it is restricted
by the dimensional reduction procedure.   In this paper we will
examine the possibility that eleven dimensional supergravity when
formulated as a coset realisation, as in reference [15], does possess a
large symmetry algebra that includes all the symmetries that occur
when it is dimensional reduced. Indeed, we conjecture that eleven
dimensional supergravity can be formulated  as a non-linear
realisation based on the group $E_{11}$. It is crucial that the
theory is formulated as   a non-linear realisation as this allows one
to treat all the fields of the theory in the same way. In
particular,   gravity and the gauge fields are all Goldstone
bosons. 
\par
We will show that the formulation of eleven
dimensional supergravity as a non-linear realisation  given in 
reference [15] is invariant under the Borel subgroup of $E_7$. We
will then argue that one can extend this formulation in such a way 
that the full theory is invariant under  a Kac-Moody algebra  and
show that if this were the case then this algebra must be  $E_{11}$.
We suggest  that this  extension can be achived in two ways. We may
take  the  local subgroup in the coset space construction to be much
larger that  the choice of the  Lorentz group in reference
[15]. We may also use an alternative description of eleven
dimensional supergravity.  In particular, we use a description of 
gravity that  possess  two fields which are
related by duality. As a result, we propose an enlarged algebra
that underlies eleven dimensional supergravity. We show that this  
algebra contains the Borel subgroup of 
$E_8$ and leads to the algebra of the ten dimensional 
IIA supergravity theory. 
\par
We will also argue that the effective action for the closed bosonic
string in twenty six dimensions is invariant under a Kac-Moody
algebra and   propose such an algebra which  has rank twenty seven. 
\par
It is likely that the proper formulation of M theory will involve
radically  new concepts. However, it is possible that these concepts
may be very difficult to guess with our current knowledge. There are,
for example, a number of indications that space-time may not be a
fundamental concept. Although non-commutative algebras have been
suggested as a replacement of space-time there does not exist any
definitive and concrete way to implement this suggestion in the
framework of M theory. The motivation for the present paper is that M theory may be
rather algebraic in character and that the known  supergravity
theories have hidden within them  
information on what are the symmetries of M theory. 
\medskip 
 {\bf {1. A Review of Kac-Moody Algebras}}
\medskip
In this paper we will need some basic facts about Kac-Moody 
algebras which we now summarise [22].  Associated with any  Kac-Moody
algebra  is a generalised Cartan matrix 
$A_{ab}$  which   satisfies the following properties:  
$$A_{aa}=2, 
\eqno(1.1)$$ 
$$A_{ab}\  {\rm for}\  a\not= b\  {\rm are\  negative\  integers\  or
\ zero}, 
\eqno(1.2)$$ 
and 
$$A_{ab}=0\  {\rm implies}\  A_{ba}=0 .
\eqno(1.3)$$ 
A   Kac-Moody  algebra can be formulated in terms of its
Chevalley generators which consist of the generators
of the  commuting Cartan  subalgebra, denoted  by 
$H_a$, as well as the generators of the positive and negative simple
roots, denoted by 
$E_a$ and  $F_a$ respectively. The Chevalley  generators are taken to
obey the Serre  relations;  
$$[H_a, H_b]= 0, 
\eqno(1.4)$$
$$[H_a, E_b]= A_{ab} E_b, 
\eqno(1.5)$$  
$$[H_a, F_b]= -A_{ab} F_b, 
\eqno(1.6)$$
$$[E_a, F_b]= \delta_{ab} H_a, 
\eqno(1.6)$$
and  
$$[E_a,\ldots [E_a, E_b]\ldots ]= 0, \  
[F_a,\ldots [F_a, F_b]\ldots]= 0
\eqno(1.7)$$ 
In equation (1.7)  there are
$1-A_{ab}$ number of 
$E_a$'s in  the first equation and the same number of $F_a$'s in the
second equation.  Given the generalised Cartan matrix 
$A_{ab}$,  one can uniquely reconstruct the entire
Kac-Moody  algebra from the above  Serre relations  by taking
the multiple  commutators of the simple root generators.  
Hence a Kac-Moody algebra is uniquely specified by its 
generalised Cartan matrix. 
\par
Any Kac-Moody  algebra is invariant under the Cartan involution
which acts on the generators as 
$$E_\alpha \to - F_\alpha,\  F_\alpha \to - E_\alpha,\ 
H_a\to -H_a
\eqno(1.8)$$
where $\alpha$ is any positive root. It is straightforward to verify
that this involution when applied to the Chevalley generators
leaves invariant the above Serre relations. We may divide the
generators of the Kac-Moody algebra into those that are even and
those that are odd under the involution.  The even generators are
given by 
$E_\alpha  - F_\alpha$.  Being  invariant under the Cartan 
involution   they must form a subgroup of
the original Kac-Moody  algebra. The remaining odd generators are
of the form  
$E_\alpha +  F_\alpha$ and $H_a$. We note that the subgroup which is
invariant under the Cartan involution contains none of the  
generators
of  the Cartan sub-algebra of the 
original Kac-Moody algebra. 
\medskip 
 {\bf {2. Identification of $E_{11}$}}
\medskip
It has been known  for many years that when 
eleven dimensional supergravity is dimensionally reduced on a torus
to $11-n$ dimensions, for
$n=1,\ldots 8$, the resulting scalars can be formulated as
a  non-linear realisations [10,11]. Let us denote the groups so
obtained by $E_n$ and the  corresponding local subgroups 
$F_n$. These are given the table below

\bigskip
\centerline {\bf Table 1 Coset Spaces of the Maximal
Supergravities}
\bigskip
$$\vbox{\tabskip=0pt \offinterlineskip

\halign to238pt{\strut#& \tabskip=1em& \ #\hfil& \ # \hfil& \ # \hfil&\ #
\tabskip=0pt\cr \noalign{\hrule}

&D&$E_n$&$F_n$&\cr \noalign{\hrule}

&11&1&1&\cr

&10, \ IIB&$SL(2)$&SO(2)&\cr

&10,\ IIA&$SO(1,1)/Z_2$&1&\cr

&9&$GL(2)$&$SO(2)$&\cr

&8&$E_3\sim SL(3)\times SL(2)$&$U(2)$&\cr

&7&$E_4\sim SL(5)$&$USp(4)$&\cr

&6&$E_5\sim SO(5,5)$&$USp(4)\times USp(4)$&\cr

&5&$E_6$&$USp(8)$&\cr

&4&$E_7$&$SU(8)$&\cr

&3&$E_8$&$S0(16)$&\cr
\noalign{\hrule}}}$$

The local subgroups  $F_n$ are the maximal compact subgroups of
$E_n$ and also have the same rank as $E_n$, namely $n$. The local
subgroup $F_n$ can be used to choose the coset representatives of
$E_n$/ $F_n$   to belong to the   
Borel sub-group of
$E_n$. We recall  that  the Borel subalgebra consists of  the 
positive root and Cartan
 generators of the original algebra and hence the group elements
of the Borel group can be written as a product  of exponentials of
these generators.   Thus the number of scalar fields is just 
the number
of  generators in the Borel subalgebra. It has been  shown that not
only the scalar sector, but the entire dimensionally reduced theory is
invariant under
$E_n$.  
\par
For $n\le 8$ the corresponding algebras found in the dimensional
reduced theory  are  finite semi-simple Lie algebras. Dimensional
reduction of eleven dimensional supergravity to  two dimensions and
one dimension is thought  to result in  theories  that are invariant  
under the affine extension of $E_8$, which is called $E_9$ or
$E_8{}^{(1)}$ and   the
hyperbolic group $E_{10}$  respectively [23]. Some evidence for this
assertion   has been given for the two dimensional case in reference
[19]. 
\par
It has not previously been proposed that eleven dimensional
supergravity itself should be invariant under any of the groups
$E_n$, however, it has been found that a number of the
SO(1,2)$\times$SO(16)  covariant objects found in the dimensional
reduced theories can be lifted to eleven dimensions. As a result, 
it has been proposed [20,26] that eleven dimensional supergravity has
some kind of  exceptional geometry. 
\par
We note that the  groups that occur in the dimensionally reduced
theories, namely $E_n$ and $F_n$, are Kac-Moody Lie groups.
Furthermore, the  local subgroups
$F_n$ for $ n=1,\ldots ,8$ are the subgroups which 
are invariant under the Cartan
involution and were 
discussed in section one. For the cases of $n=9,10$  an
involution has been introduced and used to define possible local
subgroups [24].  However, in previous discussions of the
dimensionally reduced theories,
 gravity plays a quite different role to that of the gauge fields and
scalars under the appropriate $E_n$ group.  Indeed,  the gauge
fields and scalars transform non-trivially, while gravity is  inert
under $E_n$. 
\par
When carrying out the dimensional reduction of eleven dimensional 
supergravity one has the choice whether to dualise certain fields or
leave them as they appear in the usual dimensional reduction
proceedure. The simplest example of this is  provided by a rank two
gauge fields in  four dimensions that arise from the rank three 
gauge field in eleven dimensions. These rank two gauge fields can be
dualised to become  scalars. In fact, the above
$E_n$ symmetries only emerge if one does carry out the relavent  
duality transformations.
The relationship between dualisations and the symmetries that 
emerge in
dimensionally reduced theories has been extensively studied in
reference [30]. One also has the option of introducing dual fields as
well as the original fields for all the scalars and gauge fields and
this was the path chosen in the coset formulation of this sector of
the theory in  reference [14]. This suggestion  allows the equations
of motion for the scalar and gauge  fields to be written as a
generalised self duality condition. This approach was further  
discussed in the papers of  reference [31]  where the analogy with the
Ehlers symmetry of general relativity was  explained.
The first  of these papers also includes a detailed discussion of
how the fields that occur in dimensional reduction proceedure give
rise to  the Dynkin diagram of the corresponding $E_n$ symmetries. It
was noted that if one did this for all the fields then one would
arrive at a rank eleven algebra associated with a dimensionally
reduced theory.  
\par
Let us now recall the formulation [15] of the  bosonic
sector of eleven dimensional supergravity as a non-linear
realisation. This was based on the  group ${G}_{11}$ with a Lie
algebra, also denoted,  
${\it {G}}_{11}$ whose  commutators are given by 
$$
[K^a{}_b,K^c{}_d]=\delta _b^c K^a{}_d - \delta _d^a K^c{}_b, \ 
[K^a{}_b,P_c]= - \delta _c^a P_b,\  [P_a,P_b]=0
\eqno(2.1)$$
$$
[K^a{}_b, R^{c_1\ldots c_6}]= \delta _b^{c_1}R^{ac_2\ldots c_6}+\dots, \  
 [K^a{}_b, R^{c_1\ldots c_3}]= \delta _b^{c_1}R^{a c_2 c_3}+\dots,
\eqno(2.2)$$
$$  [ R^{c_1\ldots c_3}, R^{c_4\ldots c_6}]= 2 R^{c_1\ldots c_6},\ 
[ R^{c_1\ldots c_6}, R^{d_1\ldots d_6}]=  0,
\eqno(2.3)$$
$$  [ R^{c_1\ldots c_3}, R^{a_1\ldots a_6}]= 0
\eqno(2.4)$$ 
where $+\dots $ denote the appropriate anti-symmetrisations.  
The generators 
$K^a{}_b$ and $P_c$ generate the affine group IGL(11) while  
the generators 
$R^{c_1\ldots c_3}$ and $ R^{c_1\ldots c_6}$ form a subalgebra.  
\par 
The non-linear realisation is built from group elements $g\ \in
G_{11}$ such that it is invariant under 
$$g \to g_0 g h^{-1}
\eqno(2.5)$$
where $g_0$ is a rigid element of the  group $G_{11}$ 
generated by the above 
Lie algebra 
and $h$ is a local element of the Lorentz group.  
We may take  the group element $g$ to be of the form 
$$g = e^{x^\mu P_\mu} e ^{h_{a}{}^b K^a{}_b} exp {({A_{c_1\ldots c_3} 
R^{c_1\ldots c_3}\over 3!}+ {A_{c_1\ldots c_6}
 R^{c_1\ldots c_6}\over 6!})} ,
\eqno(2.6)$$
where the fields  $h^a{}_b$, $A_{c_1\ldots c_3}$ and $A_{c_1\ldots
c_6}$  depend on $x^\mu$. 
Invariance under the rigid transformations is gained by considering
the 
$g_0$ invariant forms  given by 
$${\cal V }= g^{-1}d g -w
\eqno(2.7)$$
where $w\equiv {1\over 2} dx^\mu w_{\mu b}{}^c J^b{}_c$ is 
the Lorentz connection and so transforms as 
$$w \to h w h^{-1} +h d h^{-1}
\eqno(2.8)$$
As a result 
$${\cal V} \to h {\cal V } h^{-1}
\eqno(2.9)$$
We observe the important fact 
that the Cartan forms are inert under
the rigid transformations and only transform under the local
subgroup. 
the 
\par
Evaluating
$\cal V$ we find that it is given by 
$${\cal V}= dx^\mu(e_\mu{}^a  P_a + \Omega _a{}^b K^a{}_b
+ {1\over 3!}\tilde D_\mu A_{c_1\ldots c_3} R^{c_1\ldots c_3}
+{1\over 6!}\tilde D_\mu A_{c_1\ldots c_6} R^{c_1\ldots c_6})
\eqno(2.10)$$
where 
$$e_\mu{}^a \equiv (e^h)_\mu{}^a, \ \tilde D_\mu A_{c_1\ldots c_3}\equiv  
 \partial_{\mu} A_{c_1 c_2 c_3} 
+ ((e^{-1}\partial _\mu e)_{c_1}{}^{b}A_{b c_2 c_3}+ \dots ), 
$$
$$ 
\tilde  D_\mu A_{c_1\ldots c_6}\equiv  
\partial _\mu A_{c_1\ldots c_6}+ 
((e^{-1}\partial _\mu e)_{c_1}{}^{b}A_{b c_2 \ldots c_6}+ \dots )
- (A_{[ c_1\ldots c_3}\tilde D_\mu A_{c_4\ldots c_6]})
$$
$$\Omega_{\mu b}{}^c\equiv (e^{-1}\partial _\mu e)_b{}^c 
-w_{\mu b}{}^c,
\eqno(2.11)$$
where $+\ldots $ denotes the action of 
$(e^{-1}\partial _\mu e)$ on the other 
indices of $A_{c_1\ldots c_3}$  and $A_{c_1\ldots c_6}$. 
\par
Eleven dimensional supergravity is invariant under not just 
the group $G_{11}$, but also under the infinite dimensional group
which is the closure of $G_{11}$ and the eleven dimensional conformal
group.  This infinite dimensional group is realised by also
calculating the  Cartan forms for the conformal group and then taking
only those 
$G_{11}$ Cartan forms that can be rewritten in terms of the Cartan
forms, or other appropriate forms,  of  the conformal group. The
forms so obtained are then simultaneously covariant under both groups
and hence under the infinite dimensional group that is their closure. 
In fact,  the Cartan forms just transform under  composite 
Lorentz transformations. 
  The result of this procedure for the rank three 
and six forms is 
that one should only use the simultaneously covariant forms 
$$\tilde F_{c_1\ldots c_4}
\equiv  4(e_{[ c_1}{}^\mu \partial _\mu A_{c_2\ldots c_4]}
+ e_{[ c_1}{}^\mu 
( e^{-1} \partial_\mu  e)_{ c_2 }{\ \  }^b  
A_{b c_3 c_4]}+\ldots)
\eqno(2.12)$$
and 
$$
\tilde F_{c_1\ldots c_7}\equiv  
7(e_{[ c_1}{}^\mu(\partial _\mu A_{c_2\ldots c_7]})
+ e_{[ c_1}{}^\mu( e^{-1} \partial_\mu  e)_{ c_2 }{\ \ \ }^b A_{b
c_3 \ldots c_7]}  +\ldots  +5 \tilde F_{[c_1\ldots c_4}\tilde
F_{c_5\ldots c_7]}) 
\eqno(2.13)$$
The invariant  equation of motion is then given by 
$$
\tilde F^{c_1\ldots c_4}={1\over 7!}
\epsilon _{c_1\ldots c_{11}}\tilde F^{c_5\ldots
c_{11}}
\eqno(2.14)$$
For the Goldstone field $h_a{}^b$, that is for gravity, 
the process of
finding the simultaneously covariant forms  is more complicate 
and for this  we refer the reader to reference [15].
Although the role of the conformal group is crucial, we will for the
most part of this paper be concerned with the realisation of the
group $G_{11}$. 
It was also shown in reference [15] that the IIA supergravity theory 
can also be expressed as a non-linear realisation and it is very
likely that all  supergravity theories allow such a
description. 
\par
The scalars that occur in the dimensional reductions of eleven
dimensional supergravity have their origin in the graviton and the
rank three gauge field. Hence, if the symmetries that occur in the
dimensional reductions have their origin in eleven dimensional
supergravity, it must be in a formulation of this theory in which 
all the bosonic fields of the theory, namely the graviton and the rank
three gauge field, occur in a similar manner. One advantage of the
coset formulation of eleven dimensional supergravity is that  it does
treats gravity in the same way as the third rank gauge field; they
are both Goldstone bosons. Indeed,  general coordinate
transformations and the gauge transformations of the third rank gauge
field both arise in the  coset formulation. 
Hence, in this paper we will
search for hidden symmetries  in the coset formulation of  eleven
dimensional supergravity.  
\par
The non-linear realisation  used in reference [15] to construct the
eleven  dimensional supergravity is based on 
a finite dimensional  Lie algebra, namely $G_{11}$ which
is not  a Kac-Moody algebra.   
Furthermore, the local subgroup was 
  chosen to be just the Lorentz group and,  consequently, 
 the coset representatives  given in equation (2.6)  are 
not  members of a Borel subgroup of some larger group. 
Hence, the groups used in the coset formulation appear to be 
 unlike those  found in the dimensional reductions of eleven
dimensional supergravity. As noted above, the latter  are Kac-Moody
algebras with local  subgroups such that the coset representatives
can be  written as elements of the Borel subgroups.   From this view
point,  the coset construction of eleven dimensional supergravity 
would not  at first sight appear to  contain the Kac-Moody
groups found in the   dimensional reductions of this theory. 
\par
This strongly suggests that, although the coset formulation
of reference [15] does lead to the correct equations of motion, 
it can
be extended to  involve a    larger group than 
$G_{11}$. We will seek to achieve this in two ways.  It is
possible to   introduce a  larger group by introducing a larger local
subgroup than the   Lorentz group. This step  will not introduce extra
Goldstone bosons and only implies that the Cartan forms transform
under the larger local subgroup (see equation (2.9)). 
For example, if we adopted a larger local subgroup for 
the formulation of eleven dimensional supergravity
used in  reference [15] it  would  lead to  the same field content as
the coset formulation of reference [15], the same group element as in
equation (2.6) and the same Cartan forms as in equation (2.10). 
However, one must then show that the field equations are invariant
under the larger local subgroup in order to ensure the  invariance of
the theory under the full group.  A second possibility is to use a
coset formulation that corresponds to an alternative form of eleven
dimensional supergravity. This formulation would  describe the same
on-shell degrees of freedom, but the field content  would be 
different. As we will explain  later in this paper, gravity will be
described by two fields in a dual formulation that is analogous to
the way two fields 
$A_{a_1\ldots a_3}$ and $A_{a_1\ldots a_6}$ describe  the
non-gravitational bosonic degrees of freedom.  In fact we can carry
out both possibilities, we can first adopt a new formulation of
eleven dimensional supergravity and then enlarge the group by
introducing  a larger local subgroup. 
\par
Having carried these steps, we must take the simultaneous
non-linear realisation with the conformal group.  It is natural to
suppose that  the resulting non-linear realisation, which is
formed from two infinite dimensional groups,  is 
based on  a Kac-Moody algebra, 
$G$ and that the local subalgebra $H$ is   the subgroup  invariant
under the Cartan involution, or some modification of it.  
 The translation generator $P_a$ will play no role in
these considerations.  It would seem reasonable to
suppose that the groups
$E_n$ and $F_n$ found in the dimensional reductions of eleven
dimensional supergravity 
 are   subgroups of the
Kac-Moody Lie algebras
$G$ and $H$ respectively. 
\par
We now consider how to implement this strategy and, in particular, 
how to identify the Kac-Moody algebra $G$.   Let us  decompose the
generators   of
${{\it G}}_{11}$ into 
$${{\it G}}_{11}^+=(K^a{}_b, \ a<b,\  a,b=1,\ldots 11, \ 
R^{c_1\ldots c_3}, \  R^{c_1\ldots c_6}), 
\eqno(2.15)$$ 
$${{\it G}}_{11}^0=(H_a= K^a{}_a-K^{a+1}{}_{a+1}, a=1,\ldots 10, 
D=\sum _a  K^a{}_a)
\eqno(2.16)$$
and $K^a{}_b, \ a>b$. The generators  $K^a{}_b, \ a<b$ and $H_a$ are 
the positive root generators  and Cartan sub-algebra of SL(11)
respectively. The only other generators of SL(11) are the negative
root generators which are the generators  $K^a{}_b, \ a>b$.  As 
explained above, we   should demand that the new 
Kac-Moody Lie algebra
G  should contain, among  its positive root generators, those in 
${{\it G}}_{11}^+$ and that its Cartan subalgebra should include 
the  generators in 
${{\it G}}_{11}^0$. We define the Lie algebra which consists of 
the generators of  ${{\it G}}_{11}^+$ and ${{\it G}}_{11}^0$ 
by ${{\it G}}_{11}^{0+}$.
 \par
As a first step in identifying the Kac-Moody Lie algebra G 
we now show
that ${{\it G}}_{11}^{0+}$ contains the Borel subalgebra of $E_7$.  
We identify  the  positive root generators of $E_7$ to be just the 
generators of
${{\it G}}_{11}^+$ whose indices are restricted to  take  values
from 5 to 11. Letting this index range be denoted by 
$i,j, \ldots$,  these generators are $K^i{}_j$ for $i< j$ ,
$R^{i_1\ldots i_3}$  and $
R^{i_1\ldots i_6}$. Instead of the later generators we may
equally well use 
$$ S_i={1\over 6!}\epsilon_ {i i_1\ldots i_6}R^{i_1\ldots i_6},\  
\hat  K^i{}_j=  K^i{}_j- {1\over
7}\delta _i^j\sum_{l=5}^{11} K^l{}_l. 
\eqno(2.17)$$ 
The Cartan subalgebra generators of $E_7$ 
are identified with $H_i$ and $\hat D= \sum_{i=5}^{11} K^i{}_i $. 
Using equations (2.1) to (2.4), we find that these generators, 
together
with the generators $\hat K^i{}_j=K^i{}_j$ for $i> j$, obey the
relations
$$
[\hat K^i{}_j,\hat K^k{}_l]=\delta _j^k \hat K^i{}_l - \delta _l^i
\hat K^k{}_j, 
\eqno(2.18)$$
$$ 
 [\hat K^i{}_j, R^{k_1\ldots k_3}]= 3 \delta _j^{[k_1}R^{|i| k_2
k_3]}- {3\over 7} \delta ^i_j R^{k_1\ldots k_3} ,
\eqno(2.19)$$
$$  [ R^{i_1\ldots i_3}, R^{i_4\ldots i_6}]= 
2 \epsilon^{i_1\ldots i_6 j} S_j
\eqno(2.20)$$
$$ [ \hat D,\hat  K^i{}_j]=0,\ [ \hat D, R^{k_1\ldots k_3}]=
3R^{k_1\ldots k_3},
\eqno(2.21)$$
$$\ [\hat D, S_{k}]=6 S_{k}, 
\eqno(2.24)$$
$$[R^{i_1\ldots i_3}, S_k]=0
\eqno(2.23)$$  
and 
$$ [\hat K^i{}_j,S_k]= - \delta^i_k S_j +{1\over 7}\delta^i_j S_k,
\
\eqno(2.24)$$
\par 
In equations (2.18) to (2.24) we do indeed recognise the correct
commutators  of the Borel sub-algebra  of $E_7$ when written with
respect to its SL(7) subgroup. As proposed above, 
the positive roots generators of
$E_7$ are $\hat K^i{}_j$ for $i< j$ ,
$R^{i_1\ldots i_3}$  and $
S_{i}$ while  the Cartan
sub-algebra  generators of $E_7$ are 
$H_i= K^i{}_i-K^{i+1}{}_{i+1}$, 
$i=5,\ldots , 10$ and  $\hat D$. We note that $G_{11}$ 
also contains all the generators of the SL(7) subgroup of $E_7$ 
in the correct way. 
\par
Since ${{\it G}}_{11}$ is a symmetry of eleven dimensional
supergravity it follows that the Borel subalgebra of $E_7$ is also
a symmetry as is   the  SL(7) subgroup of $E_7$. 
\par
Under the decomposition of the adjoint (133) of $E_7$ 
into SL(7)  representations we find that  
$$133=48(K^i{}_j)+ 1(\hat D) +35(R^{i_1\ldots i_3}) 
+\bar 7 (S_i) +\bar {35}(R_{i_1\ldots i_3}) + 7 (S^i)
\eqno(2.25)$$
Thus the full $E_7$ algebra is found by adding the remaining 
 negative root generators  $S^i$  and
 $R_{i_1\ldots i_3}$. To gain a theory invariant under the full
$E_7$ we would extend the group by adding these generators to the
local subgroup and then hope to show that the equations of motion
were invariant under the now local SU(8) subgroup. In fact, the 63
of the SU(8) subgroup decomposes under SO(7) as 
$63= 21(K_{(ij)})+  \bar {35} (R_{i_1\ldots i_3}) + 7 (S^i)$. 
\par
The simple positive root generators of $E_7$ are given by 
$$E_i=K^i{}_{i+1},\ i=5,\ldots ,10,\ E_{11}=R^{91011}
\eqno(2.26)$$ 
and an 
appropriate basis for the generators of the Cartan sub-algebra is 
given by 
$$H_i= K^i{}_i-K^{i+1}{}_{i+1}, i=5,\ldots ,10, \ 
H_{11}=K^9{}_9 +K^{10}{}_{10}+K^{11}{}_{11}-{1\over 3}\hat D
\eqno(2.27)$$
It is straightforward to verify that these simple  
root  and Cartan
sub-algebra generators do indeed lead, using equations (2.18) to
(2.24),  and (1.5), to the Cartan matrix of $E_7$ that  corresponds to
its well known  Dynkin diagram. 
\par
Let us now return to the identification of the Kac-Moody  algebra G
associated with the coset formulation of eleven dimensional
supergravity. As noted in section one, it is  sufficient to determine
its  Cartan matrix. As a result, we must seek in $G_{11}$ the
Cartan subalgebra and   the simple root
generators of G. The simple positive root  generators are those from
which all positive root generators can be constructed by multiple
commutators.  It is clear that all the generators of 
${{\it G}}_{11}^+$ can be constructed, by taking repeated
commutators,  from 
$$E_a=K^a{}_{a+1}, a =1, \ldots 10, \ {\rm {and}}\ 
E_{11}= R^{91011}. 
\eqno(2.28)$$
We therefore identify these as the appropriate simple root
generators of the Kac-Moody  algebra G we seek. The  
$E_a, a =1, \ldots 10$ are the simple root generators of SL(11)   
and $E_a, a =5, \ldots 11$ are the simple root generators of
$E_7$. 
\par 
The  Cartan sub-algebra  generators of G are those
contained in
${{\it G}}_{11}^0$  in equation (2.16). Using equation (1.5),  
we may  read off the generalised Cartan matrix $A_{ij}$ of the 
Kac-Mooody algebra Lie algebra. However, in order to find an
acceptable generalised Cartan matrix $A_{ij}$, that is one that
satisfies equation (1.1)  to (1.5), we must adopt an appropriate
basis for the Cartan sub-algebra. 
\par
We observe that we
have the same number of
simple roots generators as we have generators  in the Cartan
subalgebra, namely eleven. We therefore  conclude that     we are
searching for a Kac-Moody algebra G of rank eleven which must contain
SL(11), or
$A_{10}$ and
$E_7$ as a  Lie sub-algebras. Examining their Dynkin diagrams we may
suspect that the  Kac-Moody Lie algebra we are searching for is
$E_{11}$. The    Dynkin diagram of $E_{11}$ is given in figure one
and we observe that  by deleting points in this diagram we
readily find the Dynkin diagrams of $A_{10}$ and
$E_7$. 
\par
Given the embedding of $E_7$ and $A_{10}$ it is
straightforward to show that  the only
allowed basis of the  Cartan subalgebra, in the sense of an acceptable
Cartan matrix,   is given by 
$$ H_a= K^a{}_a-K^{a+1}{}_{a+1}, a=1,\ldots ,10, \ 
H_{11}=K^9{}_9 +K^{10}{}_{10}+K^{11}{}_{11}-{1\over 3}D
\eqno(2.29)$$
These together with the simple roots $E_a, a=1,\ldots ,11$ 
  do indeed lead to the Cartan matrix of
$E_{11}$. One can also verify that the Serre relations of equation
(1.7) for the 
$E_a$ generators are satisfied, for example 
$$ [E_8, [E_8, E_{11}]]=0
\eqno(2.30)$$
\par
If we could be sure that there exists a non-linear
realisation  of eleven dimensional supergravity that  is
invariant under a  Kac-Moody Lie algebra, we could now conclude that
it is invariant under $E_{11}$. This follows from the definition of
a Kac-Moody algebra given in section one. Since the positive simple
roots exist and obey the relevant Serre relations of $E_{11}$, there
must by definition exist negative simple root generators which satisfy
all the remaining  Serre relations.  The algebra is then uniquely
specified by the occurance of    the Cartan
matrix of $E_{11}$ in these relations. 
However, we can not be sure that eleven dimensional supergravity
is invariant under  a Kac-Moody algebra. 
So far we have only shown the Serre relations of equations
(1.4), (1.5) and the part of (1.7) which involve the
$E_a$ and
$H_a$ are satisfied.  We have not shown that there exist
symmetries corresponding to the generators
$F_a$ and that these obey the remaining Serre relations. Indeed, we
can not be sure that the $E_a$ and
$H_a$ generators have multiple commutators that generate 
 the whole of
the Borel subalgebra of the Kac-Moody algebra. The problem is that 
in the absence of   the
$F_a$ generators, it can happen that a subset of  the positive
root generators $E_\alpha$ form an ideal and that this ideal may be
trivially realised in eleven dimensional supergravity. As we shall
see,  with the coset formulation of reference [15] this is indeed the
case. However,   in the next section, we propose an alternative coset
formulation of eleven dimensional supergravity which, we believe
 does realise  a Borel subgroup of a Kac-Moody algebra 
 after the full construction has been carried out. 
\par
Since $E_{11}$ is a very large algebra the problem  is more tractable
if we first  consider the smaller algebras that must arise in a
restriction of eleven dimensional supergravity. The restriction is
obtained  by  keeping  only those generators and fields with  
 indices $i,j,\ldots$
that take values $11$ to $11-n+1$. The residual invariance group one
would expect can be read off from the Dynkin diagram of
$E_{11}$  by keeping only the first $n-1$
nodes on the horizontal line in the Dynkin diagram starting from the
right as well as any nodes to which they are attached by vertical
lines. One finds that the invariance groups are the 
groups $E_n$ which are listed for  $n=1,\ldots 8$ in table one. These 
are the groups that occur in the dimensional reductions of eleven
dimensional supergravity. We would stress that the restriction
discussed here is not a dimensional reduction.  Indeed, the 
space-time is not restricted in any way and the residual groups are
sub-symmetries of eleven dimensional supergravity. 
\par
To completely specify the coset formulation of the theory 
we must specify the  local
subgroup corresponding to $E_{11}$.  It is  natural to assume
that the appropriate subgroup is just that left invariant by the
Cartan involution discussed in section one.  It is
simpler to consider the implications of this suggestion within the
context of the restriction discussed above. 
The local subgroups that
occure in the restriction must be
 the subgroups of
$E_n$ left  invariant under the
Cartan involution. As such, for $n=1,\ldots 8$, they   must be
the  subgroups 
$F_n$ that occur in table one. Clearly, $F_n$  must 
contain the local subgroup in the formulation of reference [15],
namely the Lorentz group  SO(n) for
$n=1,\ldots 10$ and SO(1,10) for n=11. 
Furthermore, $F_{n-1}$ must be  contained in
$F_n$ which suggests that $F_{n-1}$ has rank one less than $F_{n}$
and so $F_{n}$ has rank  n.  For $n=1,\ldots 8$ these conditions are
satisfied.  Although the
local subgroups specified by demanding invariance under  the Cartan
involution are also unique, for
$n\ge 9$, it is not so obvious what they are in practice. In fact,   
it could happen that a Cartan invariant subgroup of a Kac-Moody
algebra  is not itself a 
 Kac-Moody algebra.
However, assuming this not to be the case, the local subgroup for
$n=9$ is an infinite dimensional Kac-Moody algebra  of rank 9. If it
can   be obtained by adding a node to the Dynkin diagram of $F_8$,
that is 
$D_8$, and it is an affine Lie algebra 
 there are only two possibilities
$D_8{}^{(1)}$ and $B_8{}^{(1)}$ in the notation of reference [22].
 Similarly  the possible
candidates for the local subgroups  for
$n=10$  and $n=11$ can be listed. 
\par
As explained in section one, the subgroup
invariant under the Cartan involution is of the form $E_\alpha
-F_\alpha$ where $\alpha$ is a positive root. 
For   a finite dimensional semi-simple Lie  
algebras these generators are compact in the sense that if the
Cartan-Killing metric is used to evaluate the  scalar product of
these generators it is positive definite. For these groups this
definition of compactness coincides with the usual topological
definition. Hence, the local subgroups
$F_n$ for $n=1,\ldots 8$ are compact groups. However, 
the real form of the local subgroup
$F_{11}$ that occurs in  eleven dimensional supergravity must be
non-compact as it contains  SO(1,10) as a subgroup. Thus one might 
question if the subgroup $F_{11}$ defined above is really correct. 
\par
For an infinite dimensional
Kac-Moody Lie algebra one still has an analogue of a Cartan Killing
metric which when restricted to the Cartan subalgebra $H_a$ is
equal, as for the finite dimensional case, to the  Cartan matrix,
that is
$(H_a,H_b)= A_{ab}$ and  $(E_a, F_b)=\delta _{ab}$. These
formulae hold  for a simply laced algebra, but there also exists   a
suitable modification for the non-simply laced algebras. 
 For finite
dimensional Kac-Moody  algebras    the eigenvalues of the Cartan
matrix are strictly positive  and so the Cartan subgroup  generators
are compact in the above sense. However, for an infinite dimensional
Kac-Moody  algebra the eigenvalues of the Cartan matrix are not all strictly 
positive  and so the Cartan subalgebra generators are not
compact in the above sense. Indeed, in the case of 
$E_{11}$ the eigenvalues of the Cartan matrix have the signature  
$(-,+,\ldots +)$. In a similar way, the proof of the "compactness" of
the Cartan involution invariant subgroups also fails for the infinite
dimensional Kac-Moody Lie algebras and there would seem no reason 
to suppose that $F_{11}$ could not contain SO(1,10). 
\medskip 
{\bf {3.  $E_8$ Lost }}
\medskip
In section two, we observed that if a non-linear realisation 
 of eleven dimensional supergravity was based on a
Kac-Moody algebra then this would have to be $E_{11}$. However,  we
can not be sure that there does exist a non-linear realisation  that
is based on a Kac-Moody algebra. As we noted above,   if this  is not
the case, there can exist  non-trivial ideals in the algebra and
these can be trivially realised.  Nonetheless, we showed in section
two that the non-linear realisation  of reference [15] did contain
the  Borel subalgebra of $E_7$.  We have also suggested how the 
 non-linear realisation 
of eleven dimensional supergravity could be modified to possess a
full
$E_7$ symmetry.  
\par
In this section, we consider if the Borel subalgebra of $E_8$ is a
symmetry of the non-linear realisation of reference [15] of  eleven 
dimensional supergravity. Let us consider the restriction of the 
$G_{11}$ algebra such that the indices on the generators take the
values $i, j ,\dots=11,\ldots ,4$.  Introducing the generators 
$$\hat K^i{}_j=K^i{}_j- {1\over 8}\delta _i^j \sum_l K^l{}_l,\ 
\hat D=
\sum_{i=4}^{11} K^i{}_i,\   R^{i_1\ldots i_3},\  S_{k_1 k_2}= {1\over
6!}\epsilon _{k_1 k_2 i_1\ldots i_6} R^{i_1\ldots i_6}, 
\eqno(3.1)$$
we find, using the  ${\it G}_{11}$ algebra of equations (2.1) to
(2.4),  that they obey the algebra 
$$
[\hat K^i{}_j,\hat K^k{}_l]=\delta _j^k \hat K^i{}_l - \delta _l^i
\hat K^k{}_j, 
\eqno(3.2)$$
$$ 
 [\hat K^i{}_j, R^{k_1\ldots k_3}]= 3 \delta _j^{[k_1}R^{|i| \ k_2
k_3]}- {3\over 8} \delta ^i_j R^{k_1\ldots k_3} ,
\eqno(3.3)$$
$$  [ R^{i_1\ldots i_3}, R^{i_4\ldots i_6}]= 
\epsilon^{i_1\ldots i_6 j k} S_{j k}
\eqno(3.4)$$
$$ [ \hat D,\hat  K^i{}_j]=0,\ [ \hat D, R^{k_1\ldots k_3}]=
3R^{k_1\ldots k_3},\ [\hat D, S_{jk}]=6 S_{j k}
\eqno(3.5)$$ 
$$ [\hat K^i{}_j,S_{k_1 k_2}]= - 2 \delta^i_{[k_1} S_{|j| k_2]}
+{2\over 8}\delta^i_j S_{k_1 k_2},
\eqno(4.6)$$
$$ [ S_{k_1 k_2}, S_{j_1 j_2}]=0
\eqno(3.7)$$
and  
$$ [ S_{k_1 k_2}, R^{j_1\ldots j_3}]=0
\eqno(3.8)$$
\par
Let us now compare this algebra to that of the Borel subalgebra of
$E_8$. The 248 adjoint of $E_8$ decomposes into SL(8,R)
representations as $248 = 1+63 + (56+\bar {28} + \bar {8})
+ (\bar {56}+ 28 + 8)$. The $63$ are the generators of SL(8,R). The
Cartan subalgebra of $E_8$ consists of  the Cartan subalgebra  of 
SL(8,R) and the generator $1$ in the above decomposition. 
The positive
root generators of
$E_8$ are the positive root generators  of SL(8,R) as well as the
$(56+\bar {28} +\bar {8})$ and 
the negative root generators of
$E_8$ are the negative root generators  of SL(8,R) as well as the
$(\bar {56}+ 28 + 8)$. 
\par
We now wish to attempt to precisely identify the generators of the
above  restriction of
the
${\it G}_{11}$, that appear in the
algebra of equations (4.2) to (4.8),  with the Borel subalgebra
of $E_8$. In fact, it is just as  simple to also include the whole
of SL(8,R) rather than just  the Borel subalgebra of
SL(8,R). We can
identify
$K^i{}_j$ as the generators of SL(8,R),
$\hat D$ as the remaining member of the Cartan subalgebra of $E_8$,
the $ R^{k_1\ldots k_3}$ are the 56 and the $S_{k_1k_2}$ are the 
$\bar {28}$. Hence the only missing positive root generators in
the above decomposition are the 
$\bar {8}$. These form an ideal in the Borel subalgebra of $E_8$. 
In fact, the commutation relations of
equation (4.2) to (4.8) are  those of the Borel subalgebra of 
$E_8$ except that the generators  of the $\bar {8}$ are trivially
realised.  Indeed, the commutation relations of equations (4.2) to
(4.7) are precisely those of the Borel subalgebra of 
$E_8$, except that   equation (4.8)
should have a non-vanishing right-hand side that involves the missing
$\bar {8}$ generators.
\medskip 
{\bf {4.  $E_8$ Regain'd  }}
\medskip
In this section, we propose    an extension of  the algebra 
${\it G}_{11}$ that   is to be used in  a
non-linear realisation of eleven dimensional 
supergravity. Since the algebra has been enlarged the 
 formulation of eleven dimensional supergravity that results will
also be modified compared to that used in reference [15]. 
One advantage of this new algebra is that it 
  realises non-trivially the full Borel
subalgebra of
$E_8$.   This new algebra contains the generators of 
${\it G}_{11}$ as well as the  generators 
$R^{a_1\ldots a_8,b}$ which is anti-symmetric in $a_1\ldots
a_8$. The new algebra obeys   the commutators 
$$
[K^a{}_b,P_c]= - \delta _c^a P_b,\  [P_a,P_b]=0
\eqno(4.1)$$
$$
[K^a{}_b,K^c{}_d]=\delta _b^c K^a{}_d - \delta _d^a K^c{}_b,  
\eqno(4.2)$$
$$  [K^a{}_b, R^{c_1\ldots c_6}]= 
\delta _b^{c_1}R^{ac_2\ldots c_6}+\dots, \  
 [K^a{}_b, R^{c_1\ldots c_3}]= \delta _b^{c_1}R^{a c_2 c_3}+\dots,
\eqno(4.3)$$
$$[ R^{c_1\ldots c_3}, R^{c_4\ldots c_6}]= 2 R^{c_1\ldots c_6},\ 
\eqno(4.4)$$
$$
[R^{a_1\ldots a_6}, R^{b_1\ldots b_3}]
= 3  R^{a_1\ldots a_6 [b_1 b_2,b_3]}, 
\eqno(4.5)$$
$$ [ R^{a_1\ldots a_8, b} ,R^{b_1\ldots b_3}]=0, \ 
[ R^{a_1\ldots a_8, b} ,R^{b_1\ldots b_6}]=0, \ 
 [ R^{a_1\ldots a_8, b} ,R^{c_1\ldots c_8,d}]=0
\eqno(4.6)$$ 
$$ [ K^a{}_b,  R^{c_1\ldots c_8, d} ]= 
(\delta ^{c_1}_b R^{a c_2\ldots c_8, d} +\cdots) + \delta _b^d
R^{c_1\ldots c_8, a} .
\eqno(4.7)$$
Equations (2.1) to (2.3) are the same as equations (4.1) to (4.4),
but   equation (2.4) is replaced by equation (4.5) 
and it is here that the
new generator makes its appearance. 
This algebra   satisfies the Jacobi
identities provided  
$$R^{[c_1\ldots c_8, d]} =0.
\eqno(4.8)$$
\par
The restriction of this new algebra for $n=8$ gives  the generators
of equation (4.1) and in addition the generator 
$R^{i_1\ldots i_8, j} =\epsilon^{i_1\ldots i_8} S^j$. These
generators   obey the commutators of equations (4.2) to (4.7) as well
as  the relations 
$$ [\hat K^i{}_j, S^k]= \delta _j^k S^i - {1\over 8} \delta _j^i
S^k,\ 
[S_{k_1k_2}, R^{j_1j_2j_3}]=3 \delta^{ [j_1j_2}_{k_1k_2}S^{j_3]}
\eqno(4.9)$$
$$ [R^{ijk}, S^l]=0,\ [S_{ij}, S^k]=0
\eqno(4.10)$$ 
and 
$$ [\hat D, S^k]= 9 S^k. 
\eqno(4.11)$$
These equations together with those of  (4.2) to (4.7) are indeed
those of the Borel subalgebra of $E_8$ together with the remaining
generators of SL(8,R). 
\par
Thus we are led to propose that eleven dimensional supergravity can
be described by a non-linear realisation based on the algebra of
equations (4.1) to (4.7). To specify   the non-linear realisation we
must state the local subgroup. The minimal choice would be to
take just the Lorentz group.   We would then expect the 
simultaneous non-linear realisation of this coset with  the
conformal group to describe eleven dimensional supergravity.  
\par
The algebra of equations (4.2) to (4.7) is not a
Kac-Moody algebra.  However, as we explained in section two,  the
choice of local subgroup is not unique. We can further
enlarge the algebra  of equations (4.2) to (4.7) by adding generators
that belong to a local subgroup that includes the Lorentz
group and in this way hope  to arrive at a formulation of eleven
dimensional supergravity that is invariant under a Kac-Moody
algebra.   In effect one treats the generators in equations (4.2) to
(4.7), with the exception of $K^a{}_b,\  a>b$ , as part of 
the Borel sub-algebra 
of the Kac-Moody algebra and adds the corresponding  
negative root generators. 
  The final equations of motion are constructed from
the Cartan forms which transform non-trivially under the local
subgroup.  Thus a very strong constraint on the local subgroup is
that it leads to the    correct  equations of motion. 
\par
At first sight, one would not appear to arrive at $E_{11}$ through
this   procedure. However, we are required to take the closure of
the resulting algebra with the conformal algebra  and construct the
simultaneous non-linear realisation. The closure of these two
algebras will be an infinite dimensional algebra that contains a
number of the important subalgebras of $E_{11}$ and it is 
hoped that this is actually $E_{11}$. In particular, it would be
interesting to examine how the affine nature of $E_9$ arises in this 
closure. 
\par
There is also a puzzle
concerning the number of Goldstone bosons that would arise in the
construction of such a  non-linear realisation.  
It is well known that
   a non-linear realisation can  result in a  theory that has
fewer Goldstone bosons than one has generators in the coset. This is
due to a mechanism, called the inverse Higgs effect [27], that allows
one, under certain circumstances, to solve some Goldstone bosons in
terms of some of the others. In fact,  this is reason why the infinite
dimensional group which is the closure of the conformal group and the
group of affine general linear transformations leads only to the
Goldstone bosons that belong to gravity. Taking  
the group $E_{11}$ and the local subgroup that
was invariant under the Cartan involution would lead to an
infinite number of Goldstone bosons, however, we may expect
that most of these can be eliminated by the inverse Higgs effect 
to leave only a finite number of Goldstone bosons. Indeed these
residual Goldstone bosons may be just those  correspond to the
generators in  the algebra of equations (4.1) to (4.7). 
\par
We hope  to construct   this non-linear realisation   and examine
these conjectures in a future paper.  In the rest of this section, we
will examine some of the consequences of this suggestion and, as a
result, demonstrate  a number of detailed checks on the suitibility of
the algebra of equations (4.1) to (4.7). 
\par
In  a non-linear realisation based on the algebra of equations
(4.1) to (4.7),   the corresponding
Goldstone fields consist of  those of the original formulation of
reference [15], that is the fields 
$h_a{}^b,\  A_{a_1\ldots a_3},\  A_{a_1\ldots a_6}$, but 
include in addition the fields  
$A_{a_1\ldots a_8,b}$. The Goldstone bosons 
$A_{a_1\ldots a_3}$ and $A_{a_1\ldots a_6}$ lead to a first order
formulation of the equations of motion of the bosonic
non-gravitational degrees of freedom  of eleven dimensional
supergravity more usually described solely by a rank
three anti-symmetric tensor gauge field.  The field strengths of the
gauge fields  $A_{a_1\ldots a_3}$ and $ A_{a_1\ldots a_6}$ are
related  by use of the epsilon symbol. Since the only other on-shell
bosonic degrees of freedom are those of the graviton, the 
construction
implies that  there  must exist a  formulation of gravity  involving
the fields 
$h_a{}^b$ and $h_{a_1\ldots a_8,b}$. We might  expect that 
the the field strengths of these fields are related by the epsilon
symbol. Indeed, we observe that if we regard 
the lower index  on $h_a{}^b$ as that corresponding to the "gauge
field" and the other index as a type of internal index  then the
field strength would be of the form $f_{a_1a_2}{}^b$  and 
the associated
"dual gauge field" would indeed have the index structure of 
$h_{a_1\ldots a_8,}{}^b$. Hence in this construction we may expect
the graviton  equation of motion to have  a similar structure to
that of the 
 the three rank tensor gauge field.  It is to be  expected 
that   only in  such a formulation would the 
 the full symmetry of
eleven dimensional supergravity be apparent.  
\par
We now show that   such a  formulation of gravity   does 
exist.   We will  carry out the construction in a space-time of
arbitrary dimension, denoted $D$. It is well known that Einstein theory
of general relativity, whose traditional action is given by 
$$\int d^D x e R(e_\mu{}^a, w_{\mu, a}{}^b (e))  
\eqno(4.12)$$
can be rewritten in terms of the action  
$$\int d^D x e ( C_{c a,}{}^a  C^{c b,}{}_ b 
-{1\over 2}  C_{a b, c} C^{a c, b} 
-{1\over 4}  C_{a b, c} C^{a b, c}) 
\eqno(4.13)$$
where 
$$ C_{\mu\nu}{}^a= \partial _\mu e_\nu{}^a -\partial _\nu e_\mu{}^a .
\eqno(4.14)$$
It is straightforward to show that this action is equivalent to 
taking the action 
$${1\over 2} \int d^D x e (Y^{a b,c}  C_{c b,c }
+{1\over 2} Y_{a b, c} Y^{a c, b}   -{1\over 2(D-2)} 
 Y_{c a,}{}^a  Y^{c b,}{}_ b )
\eqno(4.15)$$
Introducing the field $Y_{c_1\ldots c_{(D-2)}}{}^d$ by 
$$Y^{ab,d}={1\over (D-2)!}\epsilon^{ab c_1\ldots c_{(D-2)}}
Y_{c_1\ldots c_{(D-2)}}{}^d
\eqno(4.16)$$
The  action in terms of these variables is given by 
$${1\over 2}\int d^D x (\epsilon^{\mu \nu\ldots \tau_{D-2}} 
Y_{\tau_1\ldots \tau_{D-2}}{}^d C_{\mu\nu,d}
+e(-{1\over 2} {(D-3)\over (D-2)}
Y_{\tau_1\ldots \tau_{D-2},}{}^{d} 
Y^{\tau_1\ldots \tau_{D-2},}{}_{d} 
$$
$$+{1\over 2}  (D-2) Y_{\tau_1\ldots\tau_{D-3}d,}{}^{d} 
Y^{\tau_1\ldots\tau_{D-2}e,}{}_{e}  
-{1\over 2}Y_{\tau_1\ldots\tau_{D-3}\kappa,}{}^{d}
Y^{\tau_1\ldots\tau_{D-2} d,}{}_{\kappa}))
\eqno(4.17)$$
The  equations of motion for gravity in terms of the new
variables  become 
$$ \epsilon^{\mu \tau_1\ldots \tau_{D-1}}\partial_{\tau_1} 
Y_{\tau_2\ldots \tau_{D-1}}{}^d= {\rm {terms\  of\  order}}
(Y_{\tau_1\ldots\tau_{D-2}}{}^d)^2
\eqno(4.18)$$
$$ \epsilon_{\mu\nu}{}^{\tau_1\ldots \tau_{D-2}}
Y_{\tau_1\ldots \tau_{D-2},b}= -C_{\mu\nu, b}+C_{\nu b, \mu}
-C_{\mu b, \nu}+2(e_{\nu b}C_{\mu c}{}^c -e_{\mu b}C_{\nu c}{}^c )
\eqno(4.19)$$
\par
At lowest order the first of these equations implies that 
$$ Y_{\tau_1\ldots \tau_{D-2},b}=\partial _{[ \tau_1}
h_{\tau_2\ldots \tau_{D-2}], b}. 
\eqno(4.20)$$
Consequently, we have found a description  of gravity  
constructed from the field $e_\mu^a$ and
$h_{\tau_1\ldots \tau_{D-3},}{}^b$. Taking $D=11$, this is precisely
of  the required type. 
\par
Taking a  non-linear realisation of eleven dimensional supergravity
based on the algebra of equations (4.1) to (4.7)  resolves a
puzzle concerning the relationship of the coset formulation of
IIA supergravity theory to that for the eleven dimensional theory
given in reference [15].   Although the  coset formulations 
of these two theories are
correct in the sense that  they led to  equations of
motion that do describe the degrees of freedom of both of these
theories,  
 the eleven  dimensional algebra  does not lead  in an obvious way to
the algebra of the IIA supergravity theory.  In particular, the
algebra underlying the IIA theory  involves the generators,
$K^a{}_b$ of  GL(10) as well as   the generators 
$R_{a_1\ldots a_p}$  for $p=0,1,2,3,5,6,7,8$ and it is very unclear
how the generators
$R_{a_1\ldots a_p}$ for
 $p=7,8$ can arise in   the
eleven dimensional algebra of equations (2.1) to (2.4) since this
algebra involves no generators of  rank bigger than six.  However,
we will now show that if we adopt   the new eleven dimensional algebra
of equations (4.1) to (4.7) then all the generators of the ten
dimensional IIA algebra arise naturally and obey in detail the  ten
dimensional IIA algebra of reference [15]. 
\par
Treating the eleventh index, denoted by 11, as special,    letting
$a,b = 1,\ldots ,10$ and denoting the resulting generators with a
$\tilde {\ }\ $,  the generators in the new algebra  can be written
as 
$$ \tilde K^a{}_b= K^a{}_b,\ \tilde R^a=K^a{}_{11}, \ \tilde R^{a_1
a_2}=R^{a_1 a_2 11}, \ 
\tilde R^{a_1 a_2 a_3}=R^{a_1 a_2 a_3} \ \tilde R^{a_1 \ldots
a_5}=R^{a_1\ldots a_5 11}, \
$$
$$\tilde R^{a_1 \ldots a_6}= -R^{a_1 \ldots a_6},\ 
\tilde R^{a_1 \ldots a_7}= {1\over 2} R^{a_1 \ldots a_7 11,11},\ 
$$
$$\tilde R^{a_1 \ldots a_8}= {3\over 8}R^{a_1 \ldots a_8 ,11},\ 
\tilde R= {1\over 12}(-\sum _{a=1}^{10} K^a{}_a +8 K^{11}{}_{11}) . 
\eqno(4.21)$$
Evaluating the algebra of equations (4.1)  to (4.7) for these
generators, dropping the $\tilde {\ }\ $   one indeed finds the
algebra used in the non-linear realisation of   IIA supergravity
theory of reference [15] which included the relations  
$$ [R,R^{a_1\ldots a_p}]=c_p R^{a_1\ldots a_p} , \ 
[R^{a_1\ldots a_p},R^{a_1\ldots a_q}]= c_{p,q} 
R^{a_1\ldots a_{(p+q)}}
\eqno(4.22)$$
where 
$$c_{1}=-c_{7}=-{3\over 4},\ c_{2}=-c_{6}={1\over 2},\ 
 c_{3}=-c_{5}=-{1\over 4}:
$$
$$  c_{1,2}=-c_{2,3}=-c_{3,3}=c_{2,5}= c_{1,5}=2,\ 
c_{1,7}=3, \ c_{2,6}=2,\ c_{3,5}=1
\eqno(4.23)$$
and all  other $c$'s vanish.
\par
In deriving the IIA algebra, we have required
that  the generators $R^{a_1 \ldots a_8 ,b}$ are trivally realised and so 
do not appear  in reference [15].    Use was made of   equation (4.8)
and,   as a result of this relation, the equation $-8R^{11 a_1 \ldots
a_7,a_8 }=R^{ a_1\ldots a_7a_8,11 }$. 
Clearly, this calculation provides a detailed check on the proposed
eleven dimensional  algebra. The result also implies that the IIA
algebra  contains the Borel subgroup of $E_7$ and that the
corresponding  Kac-Moody algebra as outlined in section two is
$E_{11}$.

\medskip 
{\bf { 5. The Closed Bosonic String and $K_{27}$ }}
\medskip
The closed bosonic string in 26 dimensions can also be formulated as 
 a non-linear realisation [15]. The underlying group, denoted 
$G_{26}$, has the   generators  $K^a{}_b$, $R$,  $R^{a_1a_2}$,
$R^{a_1\ldots a_{(D-4)}}$  and $R^{a_1\ldots a_{(D-2)}}$. The 
generators $K^a{}_b$ belong to the Lie algebra  GL(26) and obey
equation (2.7). Their commutation relations with the other 
generators are the analogues of equation (2.2). The remaining
commution relations  are given by 
$$ [R,R^{a_1\ldots a_p}]=c_p R^{a_1\ldots a_p} , \ 
[R^{a_1\ldots a_p},R^{a_1\ldots a_q}]= c_{p,q} R^{a_1\ldots a_{(p+q)}}
\eqno(5.1)$$
where 
$$c_{2}=-c_{D-4}={24\over (D-2)},\ 
 c_{2,D-4}=2. 
\eqno(5.2)$$
We take  all the other $c$'s to vanish and we have scaled the
generator
$R 
\to {R\over 6}$ with respect to reference [15]. The local subgroup 
is chosen to be the Lorentz group. 
\par
The non-linear realisation of 
$G_{D}$ is built out of the group element 
$g= g_h g_A$ where 
$$g_h=exp(h_a{}^b K^a{}_b)
$$
and 
$$g_A= exp ({A_{a_1\ldots a_{(D-2)}}R^{a_1\ldots a_{(D-2)}}\over (D-2)!})
exp ({A_{a_1\ldots a_{(D-4)}}R^{a_1\ldots a_{(D-4)}}\over (D-4)!})
exp ({A_{a_1a_2}R^{a_1a_2}\over (2)!})exp (AR)
\eqno(5.4)$$
We refer the reader to reference [15] for the 
derivation of the field
equations by taking the simultaneous  non-linear 
 realisation with the
conformal group. 
\par
Much of the discussion of eleven dimensional supergravity also
applies to the non-linear realisation of the effective action for the
closed bosonic string.  We can hope to extend the local subgroup and
the formulation of the effective action  such that the resulting
non-linear realisation  is invariant under a Kac-Moody algebra. In
this section, using  similar arguments as we deployed for 
eleven dimensional supergravity, we will  find what  this 
Kac-Moody algebra is likely to be. 
\par
 We
divide the generators of
${\it {G}}_{26}$ into 
$${\it {G}}_{26}^+= (K^a{}_b,\  a<b\ ,   R^{a_1a_2},
R^{a_1\ldots a_{(22)}}, R^{a_1\ldots a_{24}}), 
\eqno(5.5)$$
$${\it {G}}_{26}^0=(H_a= K^a{}_a-K^{a+1}{}_{a+1}, D=\sum_a K^a{}_a,
R) 
\eqno(5.6)$$ 
and the  remaining generators  $K^a{}_b,\  a>b$ which are
the negative root generators of SL(26). 
\par
We suppose that ${\it {G}}_{26}^+$ are 
positive root generators contained in the   Kac-Moody 
algebra  and its  Cartan subalgebra generators   contains  the
generators  in 
${\it{G}}_{26}^0$. To identify the Kac-Moody algebra we must
identify the simple roots. It is straightforward to see that 
the whole
of 
${\it {G}}_{26}^+$ can be found from multiple commutators of 
$$ K^a{}_{a+1}, R^{25 26}, R^{5\ldots 26}.
\eqno(5.7)$$
Therefore,   we identify these as the simple root generators. We
observe that we have the same number of generators in the Cartan
subalgebra as we have  simple roots, namely 27. Thus we are seeking
a Kac-Moody  algebra of rank 27. 
\par
Given these   simple roots, we must  adopt a
basis for the  Cartan subalgebra generators such that 
equation (1.5)  leads to an allowed
generalised Cartan matrix.  It turns out that the basis 
$$ H_a, \ a=1,\ldots 25,\  
H_{26}= K^{25}{}_{25}+K^{26}{}_{26}-{1\over
12}D+{1\over 6}R,\ H_{27}= K^{5}{}_{5}+\cdots +K^{26}{}_{26}-{11\over
12}D-{1\over 6}R. 
\eqno(5.8)$$
satisfies this requirement. We call this algebra $K_{27}$. The
corresponding Dynkin diagram is shown in figure two.  
We choose the local subgroup to be the one  which is
invariant under the Cartan involution. 
\par
The above process to find the Cartan matrix is not free from
ambiguity. However, we can find further evidence for the choice
$K_{27}$ by considering the restriction of the theory obtained by
taking the generators and  fields to have indices
that take values $26$ to $26-n+1$. The residual invariance group so
obtained can be read off from the Dynkin diagram of $K_{27}$ 
by keeping only the first $n-1$
nodes on the horizontal line in the Dynkin diagram starting from the
right as well as any nodes to which they are attached by vertical
lines. 
 The Dynkin diagram so
obtained is $D_{n}$ provided $3\le n \le 22$, $D_{24}$ if $n=23$
and affine $D_{24}$  if $n=24$. It is natural to take the
appropriate real form to be  O(n,n) for  $3\le n \le 22$ and 
O(24,24) if $n=23$. The corresponding local subgroups are  taken to
be those  left invariant under the Cartan involution. For these
groups, the local subgroups are then  
$O(n)\times O(n)$ for $3\le n \le 22$ and 
$O(24)\times O(24)$ if $n=23$. We denote this series of groups by
$G_n$ and their local subgroups by $H_n$. For $n<23$, 
 the coset spaces agree with that found in reference [29].
\par
We would also expect the scalar fields that occur in the
dimensional reduction  of the effective action on an n
torus to  belong to the coset space
${G_n\over K_n}$. 
This coset has   dimension 
$n^2 $ for $3\le n \le 22$ and $(n+1)^2$  if $n=23$. However, we can
calculate the number of scalar fields that occur in the
dimensional reduction.  From the graviton and anti-symmetric tensor we find
$h_i{}^j$ and $B_{ij}$, $i,j=1,\ldots n$ scalars and the vectors 
$h_\mu{}^i$ and $B_{\mu i}$. If $3\le n \le 22$ we do indeed find 
$n^2$ scalars. However, for $n=23$ the vectors  can be dualised
to form scalars and so we get $(n+1)^2$  scalars. The number of
scalars is thus in agreement with that predicted from the Dynkin
diagram of $K_{27}$. The above count does not included the scalar in
the original theory  for $n \ge 4$ and the
scalar that results from the anti-symmetric tensor $B_{\mu\nu}$ 
by dualisation in four dimensions. 
\par
It is interesting to note that the algebra $K_{27}$ contains the
algebra $E_{11}$ and one might take this as an indication that the 
closed bosonic string contains the known superstrings in ten
dimensions and in effect M theory. The closed bosonic string on a
torus is invariant under the fake monster Lie algebra and it would
be interesting to ask if $K_{27}$ was contained in this 
algebra. 
\medskip
{\bf {Conclusion}}
\medskip
In this paper we have argued that eleven dimensional supergravity 
can be formulated as a non-linear realisation based on the group
$E_{11}$. We have proposed a new formulation of eleven dimensional
supergravity that  is  an enlargement of the  algebra of
reference [15]. The simultaneous non-linear realisation of the
conformal group with this algebra, when further enlarged  by a
suitable local subalgebra, is expected to lead  to the formulation
of eleven dimensional supergravity based on $E_{11}$. 
Although much remains to be done to verify this
conjecture, it has the advantage that it requires  calculations,
that  may be lengthy, but can be carried out as a matter of
principle. 
\par
Such a non-linear realisation of eleven dimensional supergravity
would include new symmetries that mixed the degrees of freedom
described by gravity and  the rank three tensor gauge field in a very
non-trivial manner. 
\par
We have shown that some of the necessary conditions for this 
conjecture  to be
true are satisfied. In particular, 
we have  shown the presence of the
Borel subgroups of
$E_7$ and $E_8$. We have also found a formulation  of eleven
dimensional gravity based on  the fields
$h_a{}^b$ and $h_{a_1\ldots a_8}{}^a$ as required by the presence of
an  $E_8$ symmetry. 
\par
The  formulation of gravity presented in this paper,  and its origin
as a non-linear,  realisation may be of interest in its own right  in
that it may lead to the existence of new symmetries in gravity
itself. These    may  explain the presence of the Geroch and related
 symmetries [28,23] which appear so mysteriously when gravity is
constrained to possess a Killing vector. 
\par
The motivation for the current work 
was to identify some of the underlying symmetries of M theory by
studying the symmetries of the corresponding supergravity theories  
We therefore conjecture that the $E_{11}$ algebra over  an
appropriate discrete field is a symmetry of M theory and that the
closed bosonic string  has the $K_{27}$ algebra as a 
symmetry. 
\medskip
{\bf {Acknowledgment}}
\medskip
I wish to thank Matthias Gaberdiel, Neil Lambert and Andrew Pressley
for very useful discussions. 
\medskip
{\bf {References}}
\medskip
\parskip 0pt
\item{[1]} E. Cremmer, B. Julia and J. Scherk, Phys. Lett. 76B
(1978) 409.
\item{[{2}]} C. Campbell and P. West,
{\it ``$N=2$ $D=10$ nonchiral
supergravity and its spontaneous compactification.''}
Nucl.\ Phys.\ {\bf B243} (1984) 112.
\item{[{3}]} M. Huq and M. Namazie,
{\it ``Kaluza--Klein supergravity in ten dimensions''},
Class.\ Q.\ Grav.\ {\bf 2} (1985).
\item{[{4}]} F. Giani and M. Pernici,
{\it ``$N=2$ supergravity in ten dimensions''},
Phys.\ Rev.\ {\bf D30} (1984) 325.
\item{[5]} J, Schwarz and P. West,
{\it ``Symmetries and Transformation of Chiral
$N=2$ $D=10$ Supergravity''},
Phys. Lett. {\bf 126B} (1983) 301.
\item {[6]} P. Howe and P. West,
{\it ``The Complete $N=2$ $D=10$ Supergravity''},
Nucl.\ Phys.\ {\bf B238} (1984) 181.
\item {[7]} J. Schwarz,
{\it ``Covariant Field Equations of Chiral $N=2$
$D=10$ Supergravity''},
Nucl.\ Phys.\ {\bf B226} (1983) 269.
\item{[8]} L. Brink, J. Scherk and J.H. Schwarz, {\it
``Supersymmetric Yang-Mills Theories''},
Nucl. Phys. {\bf B121} (1977) 77;
F. Gliozzi, J. Scherk and D. Olive, {\it ``Supersymmetry,
Supergravity Theories and the Dual Spinor Model''}, Nucl. Phys. {\bf
B122} (1977) 253,  A.H. Chamseddine, {\it ``Interacting supergravity
in ten dimensions: the role of the six-index gauge field''},
Phys. Rev. {\bf D24} (1981) 3065;
E.\ Bergshoeff, M.\ de Roo, B.\ de Wit and P.\ van
Nieuwenhuizen, {\it ``Ten-dimensional Maxwell-Einstein
supergravity, its currents, and the issue of its auxiliary
fields''}, Nucl.\ Phys.\ {\bf B195} (1982) 97;
E.\ Bergshoeff, M.\ de Roo and B.\ de Wit, {\it ``Conformal
supergravity in ten dimensions''}, Nucl.\ Phys.\ {\bf B217} (1983)
143,  G. Chapline and N.S. Manton,
{\it ``Unification of Yang-Mills
theory and supergravity in ten dimensions''}, Phys. Lett. {\bf
120B} (1983) 105.  
{\item{[9]} S.\ Ferrara, J.\ Scherk and B.\ Zumino, 
``Algebraic Properties of Extended Supersymmetry''},
Nucl.\ Phys.\ {\bf B121} (1977) 393;
E.\ Cremmer, J.\ Scherk and S.\ Ferrara, {\it ``SU(4) Invariant
Supergravity Theory''}, Phys.\ Lett.\ {\bf 74B} (1978) 61.
\item{[10]} E. Cremmer and B. Julia,
{\it ``The $N=8$ supergravity theory. I. The Lagrangian''},
Phys.\ Lett.\ {\bf 80B} (1978) 48
\item{[11]} B.\ Julia, {\it ``Group Disintegrations''},
in {\it Superspace \&
Supergravity}, p.\ 331,  eds.\ S.W.\ Hawking  and M.\ Ro\v{c}ek,
Cambridge University Press (1981).
\item{[12]} A. Font, L. Ibanez, D. Lust and F. Quevedo, Phys. Lett.
B249 (1990) 35.
\item{[13]} C.M. Hull and P.K. Townsend,
{\it ``Unity of superstring  dualities''},
Nucl.\ Phys.\ {\bf B438} (1995) 109, hep-th/9410167.
\item{[14]} E. Cremmer, B. Julia, H. Lu and C. Pope, {\it
``Dualisation of  dualities II: Twisted self-duality of doubled fields
and superdualities"},  hep-th/9806106 
\item {[15]} P. West, {\it ``Hidden Superconformal Symmetries of M
theory"}, JHEP., hep-th/0005270. 
\item{[16]} V. Ogievetsky,  {\it ``Infinite-dimensional algebra of
general  covariance group as the closure of the finite dimensional
algebras  of conformal and linear groups"}. 
Nuovo. Cimento, 8 (1973) 988.
\item{[17]} A. Borisov and V. Ogievetsky, 
{\it ``Theory of dynamical affine and conformal 
symmetries as the theory of the gravitational field"}, 
Teor. Mat. Fiz. 21 (1974) 329. 
\item {[18]} B. de Wit and H. Nicolai, Nucl. Phys. B274 (1986)
363; H.  Nicolai Phys. Lett. 155B (1985) 47; 
\item{[19]}  H.  Nicolai, Phys. Lett. 187B (1987) 316    
\item{[20]} S. Melosch and H. Nicolai, {\it ``New Canonical
Variables for $D=11$ Supergravity"}, hep-th/9709227; 
 K. Koespell, H. Nicolai and H. Samtleben, {\it ``An Exceptional 
Geometry for $d=11$ Supergravity"}, Class. Quantum Grav. 17 (2000)
3689.
\item{[21]} G. Moore, {\it `` Finite in all directions"},
hep-th/9305139; 
 P. West, {\it ``Physical States and String Symmetries"},
hep-th/9411029,  Int.J.Mod.Phys. {\bf A10} (1995) 761. hep-th/9411029.
\item {[22]} For a review see, V. Kac, {\it ``Infinite Dimensional Lie
Algebras"}, Birkhauser, 1983. 
\item {[23]} B. Julia, in Vertex Operators in Mathematics and
Physics, Publications of the Mathematical Sciences Research
Institute no 3, Springer Verlag 1984. 
\item {[24]} H. Nicolai, {\it ``Hidden Symmetries in $d=11$
Supergravity and Beyond"}, hep-th/9906106. 
\item {[25]} E. Cremmer and B. Julia, {\it ``The SO(8) Supergravity"},
Nucl. Phys. B519 (1979) 141
\item {[26]} B. de Wit and H. Nicolai, {\it ``Hidden Symmetries ,
Central Charges and all That"}, hep-th/0011239.
\item {[27]} E. A. Ivanov and V. I. Ogievetsky, { Teor. Mat. Fiz.\/}
{\bf 25} (1975) 164. 
\item {[28]} R. Geroch, J. Math. Phys. 12 (1971) 918; 13 (1972) 394; 
H. Nicolai, {\it ``A Hyperbolic Kac-Moody Algebra from Supergravity"},
Phys. Lett. B276 (1992) 333. 
\item {[29]} J. Maharana and J. Schwarz, Nucl. Phys. B390 (1993) 3. 
\item {[30]} E. Cremmer, B. Julia, H. Lu and C. Pope, {\it
``Dualisation of  Dualities. I"},  hep-th/9710119 
\item {[31]} B. Julia, {\it "Dualities in the Classical Supergravity
Limits"}, hep-th/9805083: {\it "Superdualities: Below and beyond the
U-duality"}, hep-th/9805083.

\magnification=\magstep1

\input epsf.tex

\def\bn{\bigskip\noindent}


%

\bn\bn

\vbox{

\vskip.2cm

\hbox{\hskip.1cm\epsfbox{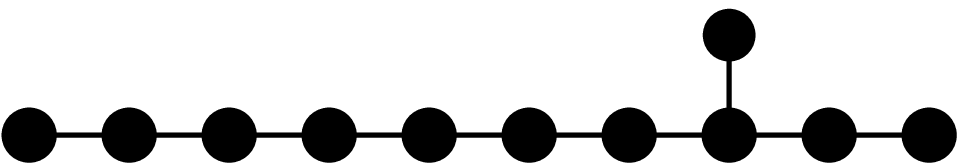}} 

\vskip1.4cm


\bn

\narrower {\bf Figure 1: }{The Dynkin Diagram of $E_{11}$. }

}

\bn

\bn\bn

\vbox{

\vskip.2cm

\hbox{\hskip.1cm\epsfbox{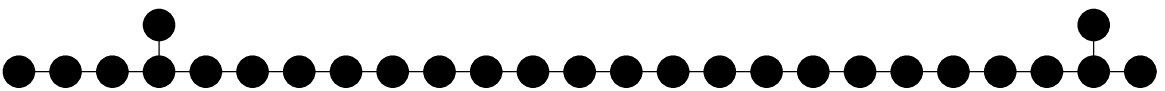}} 

\vskip1.4cm


\bn

\narrower {\bf Figure 2: }{The Dynkin Diagram of $K_{27}$.}

}

\bn

\end